\documentclass[12pt,draftclsnofoot,onecolumn]{IEEEtran}
\ifCLASSINFOpdf
\else
\fi
\usepackage{geometry}
\usepackage{graphicx}
\usepackage{amsmath}
\usepackage{cuted}
\usepackage{enumerate}
\usepackage{cite}
\usepackage{amsfonts}
\usepackage{eucal}
\usepackage{amssymb}
\usepackage{epstopdf}
\usepackage[ruled,noline]{algorithm2e}
\usepackage{color}
\usepackage{amsthm}
\usepackage{footnote}
\usepackage{ulem}
\usepackage{hyperref}
\usepackage{doi}
\usepackage{url}

\theoremstyle{definition}

\theoremstyle{remark}

\begin{document}
\title{\LARGE{A New Pricing Theory that Solves the St. Petersburg Paradox}}
\author{\IEEEauthorblockN{Dahang Li}*\footnote{\small* Dahang Li is the founder of Good Goods.}
\vspace{-2em} }

\maketitle
\begin{abstract}
\large
The St. Petersburg Paradox, an important topic in probability theory, has not been solved in the last 280 years. Since Nicolaus Bernoulli proposed the St. Petersburg Paradox in 1738, many people had tried to solve it and had proposed various explanations, but all were not satisfactory.

In this paper we propose a new pricing theory with several rules, which incidentally resolves this paradox. The new pricing theory states that so-called fair (reasonable) pricing should be judged by the seller and the buyer independently. Reasonable pricing for the seller may not be appropriate for the buyer. The seller cares about costs, while the buyer is concerned about the realistic prospect of returns.The pricing theory we proposed can be applied to financial markets to solve the confusion that financial asset return with fat tails distribution will cause the option pricing formula to fail, thus making up the theoretical defects of quantitative financial pricing theory.
\end{abstract}

\vspace{-1em}
\section{Introduction}
The St. Petersburg Paradox is still an open problem today. The infinite expectation of the St. Petersburg Paradox has been a source of contention within probability theory since its inception in early 18th century{\cite{hayden}}.

The St. Petersburg Game is that: Peter tosses a fair coin repeatedly until it shows head, he should pay $2^N$ ducats to Paul if first head appeared on the $N$-th toss. 
How much should Peter charge Paul as an entrance fee to make the game fair. 

To determine the amount of the entrance fee, we calculate Paul’s expected income. Let k be any positive integer, then the probability that the game ends at the $k$-th toss is $p(x_k)=2^{-k}$, at which time Peter will pay Paul $x_k=2^k$ ducats. 
Let $E$ denotes the expectation of Peter’s payout. Therefore 
\begin{equation}\label{expavenue}
E=\sum_{k=1}^{\infty} x_kp(x_k)=\sum_{k=1}^{\infty} 2^{k}2^{-k}=\sum_{k=1}^{\infty}1=\infty
\end{equation}
proves that no finite amount of money can be a fair entrance fee.
In short, Paul should be willing to pay an infinite price to enter this game. However, almost no rational person would agree to do that. People will only pay what they think is a moderate fee, and the fee is very limited.  In fact, it is generally no more than 20 ducats.

That is the St. Petersburg Paradox.
  \section{The St. Petersburg Game Can't Be Priced as a Limited Value}

Several approaches have been proposed for solving the paradox, such as {Expected utility theory}, Probability weighting, Rejection of mathematical expectation, Finite St. Petersburg lotteries. These views are very insightful, but none of them resolves the paradox perfectly. Moreover, these approaches are non-quantitative descriptions that cannot be applied to the actual decision-making directly.

It is generally believed that William Feller offered a mathematically correct solution involving sampling\cite{feller}. It can be understood intuitively to ``perform this game with a large number of people and calculate the expected value from the sample extraction''. In this method, when the games of infinite number of times are possible, the expectation would be infinity, and in the case of finite, the expectation will be a much smaller value. Strictly speaking, when the games are played N times, the fair price should be $\log_2 N$ .
William Feller did a good job in quantifying the fair price of the St. Petersburg Game. 

We have a different conclusion from other perspectives though.
First, take a counter-example to prove its irrationality: according to the statement, if  Peter (banker) lets Paul (player) play $1024$ games, it's fair for Peter to charge $\log_2(1024)=10$ ducats each time, and it's also fair for Paul to pay $10$ ducats each time. The problem is, if Paul plays $1024$ games at the banker Peter, and then he goes to banker Tom to play another $1024$ games. Paul should be able to make a lot of money in this way, because the game is worth 11 ducats each time for playing $2048$ times. But neither Peter nor Tom will lose. How does this work? 
Therefore, it is inappropriate that the so-called fair pricing is related to the number of times player plays.

Next, We will strictly prove that the reasonable quotation of the St Petersburg Game's banker should indeed be infinite (that is, the banker can’t quote).
Let 1 and 0 denote head and tail for the coin toss respectively. Each play of the St Petersburg Game can be represented by an infinite binary sequence, which each bit is independent, identical distribution random variable and selected from $\{0,1\}$ with equal probability. 
If the leftmost $1$ appears at the $N$-th bit (i.e. the first head appears on the $N$-th toss), Paul can get $2^N$ ducats. 
The result of the game is decided by the first N bits only.

Now, define a very simple Lottery Game $K$, where $K$ is a positive integer . The lottery number is a binary sequence with infinite digits, and each digit is independent, identical distribution random variable and selected from $\{0,1\}$ with equal probability. 
The player wins $2^K$ ducats if the first $K$ bits of the lottery number is $\overbrace{0\;\cdots\;0}^{K-1}\;1$, and nothing otherwise. 
For $E(x)=1/2^K \times 2^K=1$, the fair price of this lottery game is 1 ducat.

We take the same binary sequence as the process of the St. Petersburg Game, and as the lottery number for Lottery Game 1, Lottery Game 2, Lottery Game 3, and so on as well. Every lottery game that player plays is worth 1 ducat. Therefore, you cannot play this series of unlimited lottery games at any limited price. In addition, it is important that for the same binary sequence, the income for a single St. Petersburg Game play is equal to the total income for playing the whole lottery game set. 

Therefore, you can't play the St. Petersburg Game with a limited price. QED.
\section{The New Pricing Theory}
The St. Petersburg Paradox seems to be back to its original point. Now let's put the St. Petersburg Paradox aside and think about whether we can improve the pricing theory generally.

For a discrete event, the buyer's return is a discrete random variable $X=[x_1,x_2,x_3,\cdots]$, the probability mass function $p(x)=P(X=x),x\in X$. The expectation 
$E(X)=\sum_{i=1}^{\infty}x_i p(x_i)$.
According to the conventional pricing theory, it will be considered fair for the seller to price the event as $\mu= E(X)$ . For the continuous random variable $X= f(x)$, it is similar.
The current pricing theory will suffer from the inability to explain the St. Petersburg Paradox, and the inability to price options with certain fat tail distribution.

We propose a new pricing theory and several pricing rules to solve the issues above.

\textbf{The gist of our new pricing theory is that a fair offer to the seller is not necessarily an offer that the buyer is willing to accept. Buyers and sellers have their own decision-making mechanisms. The seller is concerned about costs, and the buyer is concerned about the realistic prospect of returns. }

\textbf{Rule 1:} For the quotation $\mu$ offered by the seller, the buyer shall judge whether the quotation is acceptable according to the following process.

At first, we define two parameters:
\begin{enumerate}
	\item Hopeless Probability $\epsilon\in [0,1]$ : let $\epsilon$ denote the probability that the buyer will ignore. More formally, the buyer ignores the possibility that the probability of some opportunities is not greater than $\epsilon$, and does not pay for such opportunities.
	\item Cost-effectiveness Factor $k$ :	let k denote the buyer’s investment preference. k=1 means that the buyer seeks fair deals, and $k<1$ means that the buyer seeks stable profit opportunities (e.g. the wrong pricing of the seller), and $k>1$ means that the buyer is speculating (e.g. gaming).
\end{enumerate}

Note that every buyer can choose his own $\epsilon$ and $k$ values.
Suppose that $N_\epsilon$ is the minimum positive integer such that
\begin{equation}\label{eq2}
 \sum_{i=N_{\epsilon}+1}^{\infty}p(x_i)\leq \epsilon
\end{equation}

Since 
\begin{equation}\label{eq3}
  \sum\limits_{i=1}^{N_\epsilon}p(x_i)+\sum_{i=N_{\epsilon}+1}^{\infty}p(x_i)=\sum_{i=1}^{\infty}p(x_i)=1
\end{equation}
we have
\begin{equation}\label{eq4}
 \sum\limits_{i=1}^{N_\epsilon}p(x_i)=1-\sum_{i=N_{\epsilon}+1}^{\infty}p(x_i)\geq 1-\epsilon
\end{equation}
and let
\begin{equation}\label{eq5}
E_{\epsilon}= \sum\limits_{i=1}^{N_\epsilon}x_ip(x_i)
\end{equation}
then $\mu$ is an acceptable price for the buyer if $\mu \leq kE_\epsilon$.

\textbf{Rule 2:} In the case that the contract must be executed at the end, the seller determines the quotation $\mu$ with the expectation just like traditional pricing theory. That is

\begin{equation}\label{eq6}
\mu\geq E(X)=\sum\limits_{i=1}^{\infty}x_ip(x_i)
\end{equation}

\textbf{Rule 3:} In the case that the seller can close the contract during the process (such as stock option), the seller uses a similar method as the buyer in Rule 1 to determining the quotation $\mu$.
Thus, we have
\begin{equation}\label{eq7}
\epsilon\in [0,1] \stackrel{\eqref{eq2}}{\longrightarrow}N_{\epsilon}\stackrel{\eqref{eq5}}{\longrightarrow} E_{\epsilon}
\end{equation}
and let $\mu=kE_\epsilon,k\geq 1$ for gaining profit.
When $\epsilon=0$ and $k=1$, we have $\mu=E$, this is the conventional fair price.
When $\epsilon>0$, the seller should choose $k > 1$, since he bears the additional risk of $\epsilon$ without calculating the cost for gambling.
The premise that sellers can use Rule 3 is that the exchange has a margin system for risk control.
\section{Resolve the St. Petersburg Paradox}
Based on the pricing rules proposed in section III, the seller's pricing should be infinite for the St. Petersburg Game. That is, the seller cannot provide appropriate price for the game to the buyer. This conclusion has been proved in section II, but it does not eliminate most people's doubts. 

Camerer\cite{camerer}, Hayden\cite{hayden} and many authors have discussed how much the St. Petersburg Game is worth paying for. 
Usually, the values are small, and almost do not exceed 20 ducats. The Rule 1 can explain this phenomenon well.

Let's first look at the American Powerball as a comparison, with a bet of two dollars, and the odds of winning the jackpot is one in 292.2 million. In 2019, the Jackpot Amount averaged about 171 million. Many people are willing to bet.

Back to the St. Petersburg Game, for $2^{-28}=\frac{1}{268435456} \approx \frac{1}{292200000}$, let $\epsilon = \frac{1}{2^{28}}$, we have
\begin{equation}\label{eq8}
  \frac{1}{2^{28}}=\epsilon \geq \sum\limits_{i=N_{\epsilon}+1}^{\infty}p(x_i)=\sum\limits_{i=N_{\epsilon}+1}^{\infty}2^{-i}=2^{-N_{\epsilon}}
\end{equation}

Hence, $N_{\epsilon}=28$, we will omit those bonuses of $2^{29}$,$2^{30}$,$2^{31}$,$\cdots$
\begin{equation}\label{eq9}
\begin{split}
  E_{\epsilon}= \sum\limits_{i=1}^{N_\epsilon}x_ip(x_i)=28
\end{split}
\end{equation}

In other words, it is appropriate for a rational player to spend 28 ducats to play the St. Petersburg Game. However, almost no one is willing to spend 28 ducats to play the St. Petersburg Game. Why?
\begin{equation}\label{eq10}
\begin{split}
  p(X_{N_{\epsilon}})=2^{-28}=\frac{1}{268435456}>\frac{1}{292200000}\\
  X_{N_{\epsilon}}=2^{28}=268435456>1.71\times {10^8}\\
\end{split}
\end{equation}

An important difference between the Powerball lottery and the St. Petersburg Game is that the St. Petersburg Game costs a lot more with close odds and amount of the jackpot. Powerball lottery only costs 2 (dollars), while St. Petersburg Game costs 28 (ducats). 

In this regard, the St. Petersburg Game is poorly designed, with a large bet amount, everyone wins, and has not concentrated the cost on the players' attention. This makes people set the Cost-effectiveness Factor k to a very low value while playing the St. Petersburg Game, such as $k=0.5$; on the contrary, they will accept a much more large k even if $k>1$ while buying the Powerball lottery.

Sellers price based on cost, but buyers are not necessarily willing to pay for all costs. 
That is the key to the St. Petersburg Paradox.
\section{Fat Tail and Option Pricing}
The correct description of asset return distribution is directly related to the correctness of portfolio selection, the effectiveness of risk management and the rationality of option pricing. In the classical efficient market hypothesis, stock returns are usually assumed to follow a normal distribution in which the ends of the distribution are thin tails. However, in reality, markets do not behave in this way, and financial asset returns do not simply follow geometric Brownian motion. The extreme events similar to the financial crisis occur much more frequently than expected, and the empirical distribution of returns has obvious fat tails.
Mandelbrot\cite{mandelbrot}, Fama\cite{fama}, and Kon\cite{kon} all report evidence that stock returns are not consistent with the random walk theory.

There are two wonderful passages in Taleb's book\cite{taleb}:"There is an example academic literature trying to maintain us that options are not rational to own because some options are overpriced, and they are embedded overpriced according to business school methods of computing risks that do not take into account the possibility of rare events" and "Further, casino bets and lottery tickets also have a known maximum upside – in real life, the sky is often the limit, and the difference between the two cases can be significant. Risk taking ain't gambling, and optionality ain't lottery tickets".
Coincidentally, our paper considers lottery and options together.

It is of great importance to accurately describe the statistical characteristics of the financial asset return distributions, which is the premise of the correct option pricing. Finance engineers often choose the Lévy Distribution to model price changes in markets. The fat tail or slow fall off indicates that this distribution model is a good match for what happens after prices change. The normal (Gaussian) distribution, in which $\alpha=2$, is a special case of a Lévy distribution. One case of a non-Gaussian Lévy distribution is the Cauchy distribution in which $\alpha=1$.

The Lévy Stable Distribution ($1\leq \alpha <2$) implies infinite variance. Having an infinite variance does not prevent a distribution from becoming proper, but it does make it quite peculiar. 
In standard financial theory, volatility ($\sigma$) is the most important parameter and is usually measured by the standard deviation of financial asset returns. Volatility is used in the financial calculation of risk and option pricing. 
The variance of Lévy Stable Distribution ($\sigma^2=\infty$) will make the option pricing formula yield a meaningless answer. 
Conventional financial theories have been challenged for their lack of capability of realistically explaining those meaningless results.

For example, for Cauchy distribution, its density function is
\begin{equation}\label{eq11}
 p(x)=\frac{1}{\pi (x^2+1)}
\end{equation}

Calculating the European call option price $C$ in which $S(T)$ is the price of the stock at time $T$, $K$ is the exercise price, and the risk-free rate of return is $r$, the following equation is applicable: 
 \begin{equation}
\begin{split}
  C&=e^{-rT}E[\max(S(T)-K,0)]\\
&=e^{-rT}\int_{K}^{\infty}(x-K)p(x)dx=e^{-rT}\int_{K}^{\infty}(x-K)\frac{1}{\pi(x^2+1)}dx\\
&=e^{-rT}\left(\int_{K}^{\infty}\frac{x}{\pi(x^2+1)}dx-K\int_{K}^{\infty}\frac{1}{\pi(x^2+1)}dx\right)>e^{-rT}\left(\int_K^{\infty}\frac{x}{\pi (x^2+1)}dx-K\right)\\
&\geq e^{-rT}\left(\int_{\max(K,2)}^{\infty}\frac{x}{\pi(x^2+1)}dx-K\right)\geq e^{-rT}\left(\int_{\max(K,2)}^{\infty}\frac{x}{\pi(x^2+x)}dx-K\right)\\
&= e^{-rT}\left(\int_{\max(K,2)}^{\infty}\frac{1}{\pi(x+1)}dx-K\right) = e^{-rT}\left(\frac{\ln(x+1)}{\pi}|_{\max(K,2)}^{\infty}-K\right)=\infty\\
\end{split} 
\end{equation}

This equation presents a problem.

Now, in terms of option pricing, according to the new pricing theory (rules) proposed in section III, $E[\max(S(T)-K,0)]$ and $E[\max(K-S(T),0)]$ no longer have to be finite.

When calculating the option price, a sufficiently small $\epsilon[0,1]$ should be selected as the hopeless probability, and we assume that $U$ satisfies the equation:
\begin{equation}\label{eq13}
    \int_{U}^{\infty}p(x)dX\leq \epsilon
\end{equation}
then 
\begin{equation}
C= e^{-rT}\int_{-\infty}^{U}\max(x-K,0)p(x)dx=e^{-rT}\int_{K}^{U}(x-K)p(x)dx
\end{equation}

Considering the actual situation of the real securities market, we can consider that $S_T<100S$ is absolutely true, that is, $U=100S$ fulfills \eqref{eq13}. 

Thus
\begin{equation}\label{eq14}
 C = e^{-rT}\int_{K}^{100S}(x-K)p(x)dx
\end{equation}

Similarly, we have 
\begin{equation}\label{eq15}
 P=e^{-rT}\int_{0.01S}^K(K-x)p(x)dx
\end{equation}

Finally, we can conclude that there is no need to worry about the probability density distribution causing the option pricing formula to fail. The upper and lower limits of the integration operation can be truncated.
\section{Conclusions}
In this paper, we proposed a new pricing theory. An offer that is fair to the seller is not necessarily an offer that the buyer is willing to accept. The buyer and the seller have their own decision-making mechanisms. The seller is concerned about costs, and the buyer is concerned about the realistic prospect of returns. We also proposed three specific quantitative pricing rules. Moreover, our paper solves the St. Petersburg Paradox perfectly and gives a method for pricing options of financial assets with fat tail distribution.

For the next step, there will be good prospects for studying the distribution model (with fat tails) of actual stock/index and the method of determining the Hopeless Probability $\epsilon$.

\bibliographystyle{IEEEtran}
\bibliography{References}
\end{document}